\documentclass[twocolumn]{aastex63}
\usepackage[normalem]{ulem}  
\usepackage[T1]{fontenc}
\usepackage{ae,aecompl}
\usepackage{graphicx}
\usepackage{amsmath}
\usepackage{amssymb}
\usepackage{supertabular}
\usepackage{hyperref}
\usepackage{xcolor}
\usepackage{academicons}

\def\bea{\begin{eqnarray}} \def\eea{\end{eqnarray}}
\def\fm3{\;\text{fm}^{-3}}

\newcommand{\Msun}{\,M_{\odot}}

\newcommand{\D}{{\rm d}}

\begin{document}

\title{Dark matter admixed neutron star properties in the light of X-ray pulse profile observations}

\author{Zhiqiang Miao} 
\affiliation{Department of Astronomy, Xiamen University, Xiamen, Fujian 361005, China; liang@xmu.edu.cn; fenghuang@xmu.edu.cn}

\author{Yaofeng Zhu}
\affiliation{Department of Astronomy, Xiamen University, Xiamen, Fujian 361005, China; liang@xmu.edu.cn; fenghuang@xmu.edu.cn}

\author{Ang Li}
\affiliation{Department of Astronomy, Xiamen University, Xiamen, Fujian 361005, China; liang@xmu.edu.cn; fenghuang@xmu.edu.cn}

\author{Feng Huang}
\affiliation{Department of Astronomy, Xiamen University, Xiamen, Fujian 361005, China; liang@xmu.edu.cn; fenghuang@xmu.edu.cn}


\begin{abstract}

The distribution of the dark matter (DM) in DM-admixed-neutron stars (DANSs) is supposed to be either a dense dark core or an extended dark halo, which is subject to the DM fraction of DANS ($f_{\chi}$) and the DM properties, such as the mass ($m_{\chi}$) and the strength of the self-interaction ($y$). In this paper, we perform an in-depth analysis of the formation criterion for dark core/dark halo and point out that the relative distribution of these two components is essentially determined by the ratio of the central enthalpy of the DM component to that of the baryonic matter component inside DANSs. For the critical case where the radii of DM and baryonic matter are the same, we further derive an analytical formula to describe the dependence of $f^{\rm crit}_{\chi}$ on $m_{\chi}$ and $y$ for given DANS mass. The relative distribution of the two components in DANSs can lead to different observational effects. We here focus on the modification of the pulsar pulse profile due to the extra light-bending effect in the case of a dark-halo existence and conduct the first investigation of the dark-halo effects on the pulse profile. We find that the peak flux deviation is strongly dependent on the ratio of the halo mass to the radius of the DM component. Lastly, we perform Bayesian parameter estimation on the DM particle properties based on the recent X-ray observations of PSR J0030+0451 and PSR J0740+6620 by the Neutron Star Interior Composition Explorer.

\end{abstract}


\section{Introduction} 

Neutron stars (NSs) could act as astrophysical laboratories to test the possible effects of dark matter (DM) and indirectly measure DM particle properties~\citep{1989PhRvD..40.3221G}.
Multimessenger observations have been explored to constrain the structure of the DM-admixed NSs (DANSs), which in turn puts limits on DM parameter space. 
In this context, there are many studies on how DM may impact various observable properties of NSs, such as the mass-radius relation~\citep{2006PhRvD..74f3003N,2009APh....32..278S,2011PhLB..695...19C,2011PhRvD..84j7301L,2012APh....37...70L,2012JCAP...10..031L,2012PhRvD..85j3528L,2013PhLB..725..200G,2014PhRvC..89b5803X,2015PhRvD..92f3526K,2015APh....62..115L,2017ApJ...835...33R,2017EPJC...77..440M,2017PhRvD..96h3004P,2018JPhG...45eLT01M,2019IJMPD..2850148W}, the surface temperature~\citep{2008PhRvD..77d3515B,2008PhRvD..77b3006K,2010A&A...522A..16G,2010PhRvD..81l3521D,2010PhRvD..82f3531K,2011PhRvD..84j3510F}, the kinematics and rotation properties~\citep{2012PhLB..711....6P}, the tidal deformability~\citep{2018PhRvD..97l3007E,2019JCAP...07..012N,2019PhRvD..99d3016D,2020PhRvD.102b3025F,2020MNRAS.495.4893D,2020JPhG...47i5202Q,2021PhRvD.104f3028D,2021JCAP...10..086H,2021ApJ...922..242L,2021arXiv211113289S,2021MNRAS.504.3354S,2022PhRvD.105d3013D,2022PhRvD.105f3005D,2022PhRvD.105b3001R}, or the gravitational wave waveform during the postmerger stage~\citep{2018JHEP...11..096K,2019PhRvD.100d4049B,2020arXiv201211908B}.

In general, DM can form either a dense core inside the NS or an extended halo surrounding the NS, which would result in different or even opposite effects on some properties of DANSs. 
Since the detection of the GW170817 binary NS merger~\citep{2017PhRvL.119p1101A}, the DANS study through the tidal deformability has gained much attention.
For example, it is found that an extended dark halo could dramatically enhance the stellar tidal deformability $\Lambda$~\citep{2019JCAP...07..012N}, while a dense dark core will affect it in an opposite way~\citep{2018PhRvD..97l3007E,2021arXiv211113289S}. 
The upper bound of tidal deformability for a NS with $1.4\Msun$ $\Lambda_{1.4}\leq800$~\citep{2017PhRvL.119p1101A} reported by the LIGO/Virgo collaboration has been used to provide new constraints on the DM properties~\citep{2018PhRvD..97l3007E,2020JPhG...47i5202Q,2021JCAP...10..086H,2021arXiv211113289S}. 
It is suggested that future detection of large tidal deformability, which exceeds the expected range of NS, could be interpreted as a dark halo~\citep{2019JCAP...07..012N}.
The possibility of the $2.6\,M_\odot$ compact object in the binary merger GW190814~\citep{2020ApJ...896L..44A} being a DANS has also been investigated in the literature~\citep{2021PhRvD.104f3028D,2021ApJ...922..242L,2021arXiv211012972W}. 

The formation of a dark halo or dark core should be determined by the nature of DM and the amount of DM accumulated in DANSs. 
As illustrated in e.g., \citet{2020PhRvD.102f3028I,2022PhRvD.105b3001R}, light DM particles tend to form a dark halo while heavy ones prefer to form a dark core. 
In the scenario of self-interacting DM, the strength of DM self-interaction also plays an important role here. Qualitatively, the larger the strength of the self-repulsive interaction, the easier it is for DM to occupy a larger volume; that is, the easier it is to form a dark halo. 
However, there is no general criterion in the current literature to determine whether it is a DM halo or a dark core when a specific DM model is assumed. 
This paper aims to perform an in-depth analysis of the formation criterion for dark core/dark halo. As deduced in later sections, we point out that the central enthalpy is the key factor. 
For example, if the central enthalpy of baryonic matter~($h^c_{B}$) is larger than that of DM~($h^c_{D}$), then there forms a dark core; otherwise, there should be a dark halo. 
For the critical case where the radii of DM and baryonic matter are the same, the amount of DM in DANS, represented by the DM fraction of DANS ($f_{\chi}$) at a fixed DANS total mass~($M_T$), should depend on a specific DM model. 
We further derive an analytical formula to describe the dependence of $f^{\rm crit}_{\chi}$ on the DM particle mass ($m_{\chi}$) and the strength of DM self-interaction.  

X-ray oscillations have been observed from a few millisecond pulsars. The shape of the pulsation in the X-ray flux (i.e. pulse profile) encodes information about the surface properties of the NS as well as its exterior spacetime.
Therefore, when a dark halo formed in DANSs, pulses emitted from the surface of baryonic matter should suffer an additional gravitational potential and modification of the pulse profile is expected.
\cite{2022PhRvD.105f3005D} firstly mentioned such kind of effect in their discussion, but so far, no detailed numerical calculations have been performed on how the dark halo would modify the pulse profile emitted from the surface of the baryon component. 
Such an analysis is carried out for the first time in the present work.
As we will see later, this modification sensitively depends on the ratio of the halo mass to the radius of DM component $M_{\rm halo}/R_D$. 
For compact dark halo with $M_{\rm halo}/R_D\simeq0.01$, the modification in pulse profile can reach about 10\%. 
Therefore modified pulse-profile modeling is necessary when one applies the NICER measurements to study DANSs.
As an example, we show how we incorporate the recent NICER data for studying DANSs and constraining the dark DM particle properties. 

This paper is organized as follows. In Section~\ref{sec:dark matter EOS} we present the formalism of self-interacting fermionic DM. In Section~\ref{sec:DANS structure} we study the structure of DANSs. In particular, we focus on deriving the criterion of the formation of core/halo and estimating the critical amount of DM. In Section~\ref{sec:pulse profile} we discuss the effect of dark halos on the DANS pulse profiles. Then in Section~\ref{sec:dark matter properties} we perform bayesian parameter estimation on DM properties by using the NICER results from X-ray pulse-profile modeling. We summarize in Section~\ref{sec:summary}.

\section{Dark matter equation of state}\label{sec:dark matter EOS}

In this work, we consider DM to be made of self-interacting fermions with masses from MeV to GeV. 
We assume that the DM particles are non-annihilating as in asymmetric DM~\citep{2014PhR...537...91Z}. 
Neglecting finite temperature effects, the energy density of DM with particle mass $m_\chi$ is given by~\citep{2006PhRvD..74f3003N,2019JCAP...07..012N}: 
\begin{equation}\label{eq:gene_eos}
    \epsilon_\chi = \epsilon_{\rm kin} +m_\chi n_\chi+\frac{\left(\hbar c\right)^3n_\chi^2}{m_I^2}\ ,
\end{equation}
where $n_\chi$ is the DM number density and $\epsilon_{\rm kin}$ is the DM kinetic energy. $m_I$ represents the energy scale of the self-interaction via a Yukawa type in the repulsive case~\citep[see more discussions in][]{2015PhRvD..92f3526K,2019PhRvD..99h3008G}, providing stabilization with respect to the gravitational collapse~\citep{2019JCAP...07..012N,2020PhRvD.102f3028I}.
For weak interaction, $m_I\sim300\,{\rm GeV}$ as mediated by W or Z bosons, while for strongly interacting DM particles, $m_I$ is assumed to be $\sim100\,{\rm MeV}$ according to chiral perturbation theory~\citep{2006PhRvD..74f3003N}.

The kinetic energy can be calculated by the model of an ideal Fermi gas: 
\begin{equation}
    \epsilon_{\rm kin} = \frac{1}{\pi^2\left(\hbar c\right)^3}\int_{0}^{k_F}dk k^2(\sqrt{k^2+m_\chi^2}-m_\chi)\ ,
\end{equation}
where $k_F = \hbar c(3\pi^2n_\chi)^{1/3}$ is the Fermi momentum. 
By denoting $x \equiv k_F/m_\chi$ and $y \equiv m_\chi/m_I$, 
the energy density and pressure are determined as:
\begin{equation}\label{eq:e_chi}
      \epsilon_\chi = \frac{m_\chi^4}{8\pi^2(\hbar c)^3}\left[(2x^3+x)\sqrt{1+x^2}-\sinh^{-1}(x)\right] 
    + \frac{m_\chi^4y^2x^6 }{(3\pi^2)^2(\hbar c)^3} \ ,
\end{equation}

\begin{equation}
    P_\chi = \frac{m_\chi^4}{24\pi^2(\hbar c)^3}\left[(2x^3-3x)\sqrt{1+x^2}+3\sinh^{-1}(x)\right]
    + \frac{m_\chi^4y^2x^6}{(3\pi^2)^2(\hbar c)^3}\ .     
\end{equation}
The DM equation of state (EOS), i.e., the pressure-density relation, is then obtained.
 
\section{Dark matter admixed neutron star structure}\label{sec:DANS structure}

\begin{figure}
\centering
\includegraphics[width=3.4in]{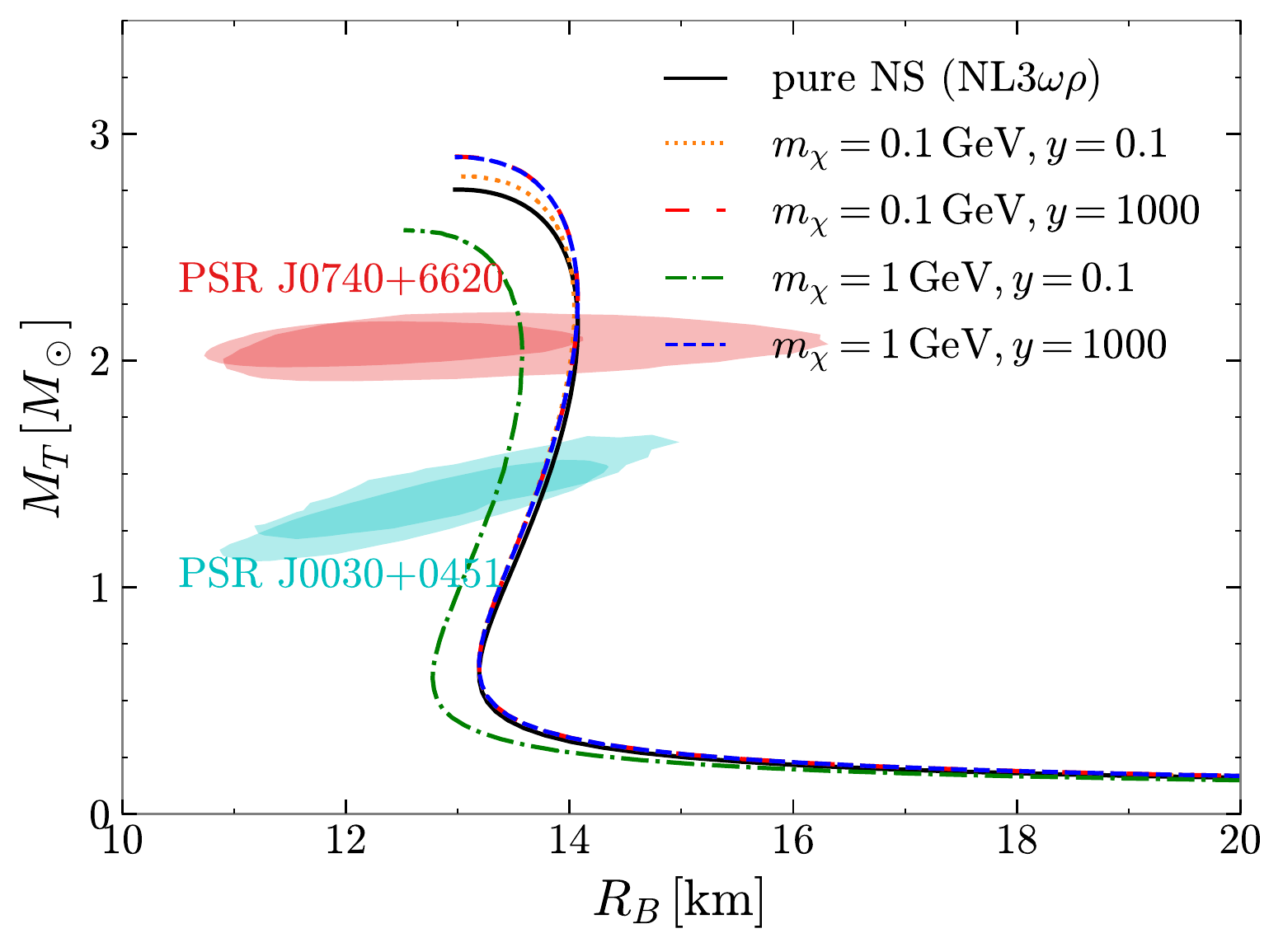}
\caption{Total gravitational mass as a function of the radius of the baryon component. The black solid curve is for pure NSs from the NL3$\omega\rho$ EOS~\citep{2001PhRvL..86.5647H} without DM, whereas the other curves are for DANSs with DM fraction $f_\chi=0.05$~\citep{2018PhRvD..97l3007E}. 
The NICER's mass-radius measurements for PSR J0030+0451~\citep{2019ApJ...887L..24M,2019ApJ...887L..21R} and PSR J0740+6620~\citep{2021ApJ...918L..27R,2021ApJ...918L..28M} from X-ray pulse profile modeling are also shown in the shaded regions, respectively. 
}
\label{fig:m-r relation}
\end{figure}

The DANS stable configuration in hydrostatic equilibrium is obtained by solving the two-fluid Tolman-Oppenheimer-Volkoff (TOV) equations~\citep{1972PThPh..47..444K,2009APh....32..278S,2011PhLB..695...19C} for the pressure and the enclosed mass,
\begin{align}
    \frac{\D P_B}{\D r}=&-\frac{GM(r)\epsilon_B(r)}{r^2}\nonumber\\
    &\times\left(1+\frac{4\pi r^3P(r)}{M(r)}\right) \left(1+\frac{P_B(r)}{\epsilon_B(r)}\right)\left(1-\frac{2GM(r)}{r}\right)^{-1}\label{eq:tov1} ,\\  
    \frac{\D P_\chi}{\D r}=&-\frac{GM(r)\epsilon_\chi(r)}{r^2}\nonumber\\
    &\times\left(1+\frac{4\pi r^3P(r)}{M(r)}\right)\left(1+\frac{P_\chi(r)}{\epsilon_\chi(r)}\right)\left(1-\frac{2GM(r)}{r}\right)^{-1}\label{eq:tov2} ,\\
    \frac{\D M_B}{\D r} = &4\pi r^2\epsilon_B(r)\label{eq:tov3} ,\\
    \frac{\D M_D}{\D r} = &4\pi r^2\epsilon_\chi(r)\label{eq:tov4},
\end{align}
where $M(r)=M_B(r)+M_D(r)$ and $P(r)=P_B(r)+P_D(r)$ with the subscript index `B' and `D' stand for the baryonic matter and DM components respectively. 
Given the central pressures of the two components $P_{B,c}$ and $P_{\chi,c}$ as initial conditions, we solve Eqs.~(\ref{eq:tov1}-\ref{eq:tov4}) by employing the Runge-Kutta 4th order method. In each iteration step we calculate the densities of the two components from their pressures using corresponding EOSs. We integrate Eqs.~(\ref{eq:tov1}-\ref{eq:tov4}) outward until the pressures of the baryonic matter and DM vanish.
This gives the radii of the baryon component $R_B$ and the DM component $R_D$. 
The gravitational masses of two components are then $M_B(R_B)$ and $M_D(R_D)$, respectively. 
Hereafter, we denote $f_\chi = M_D(R_D)/M_T$ as the DM fraction, where $M_T = M_B(R_B)+M_D(R_D)$ is the total gravitational mass. 
For the description of the baryonic matter, we use the NL3$\omega\rho$ EOS, developed in the relativistic effective field theories to fit nuclear matter saturation and properties of finite nuclei~\citep{2001PhRvL..86.5647H}.
The DM EOS can be varied by changing the particle mass $m_\chi$ and the interaction strength $y$.
Note that we have assumed the baryonic matter and DM components couple essentially only through gravity since the non-gravitational interactions between them could be negligibly small~\citep{2016JPhG...43a3001M,2016PhRvC..94d5805R}. 

\subsection{Mass-radius relation}\label{subsec:M-R}

\begin{figure}
\centering
\includegraphics[width=3.4in]{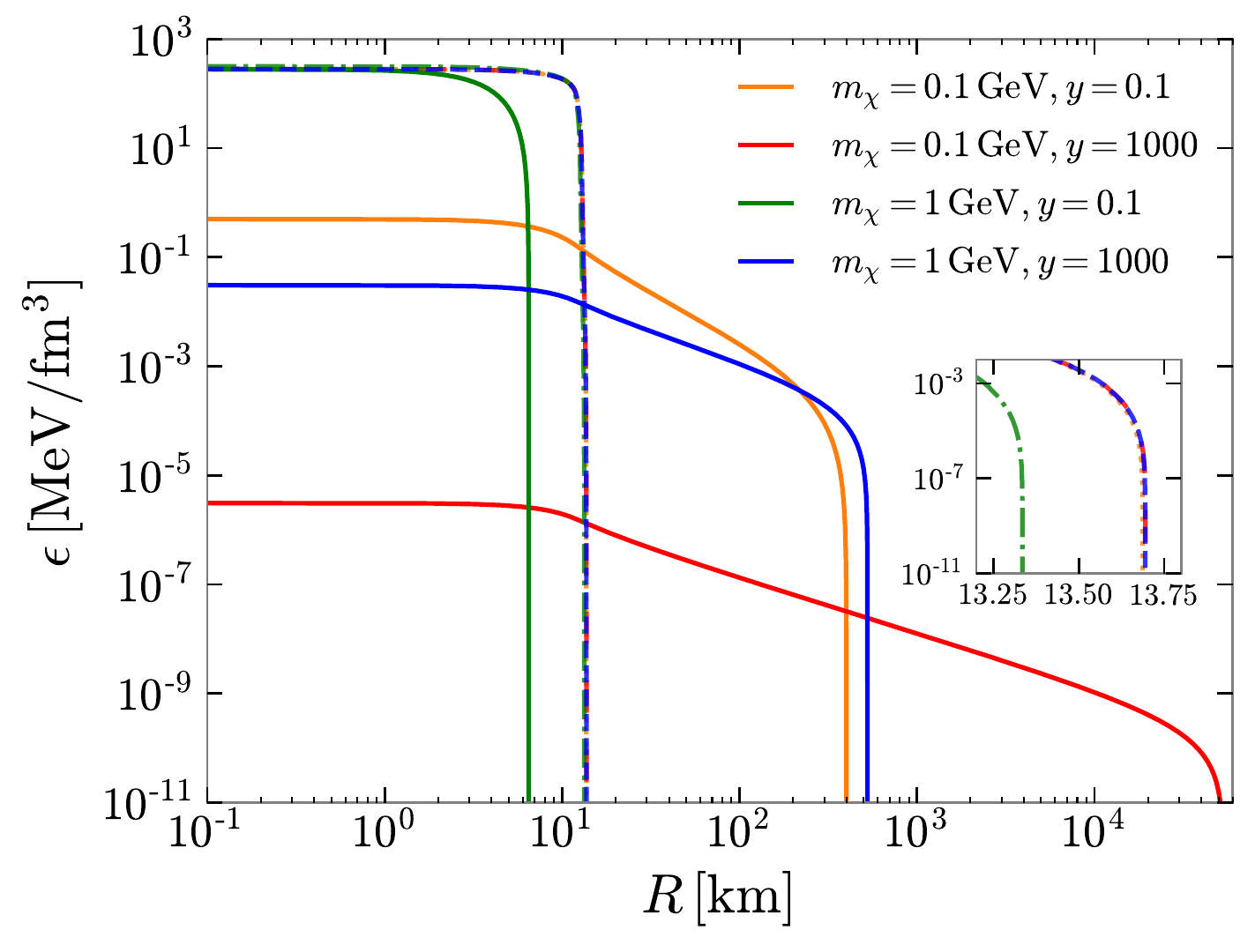}
\caption{Energy density as a function of the stellar radius for $M_T=1.4\,\Msun$ DANSs.
The DM fraction is fixed at $f_\chi=0.05$~\citep{2018PhRvD..97l3007E}. 
Solid curves show density profiles for the DM component, while dashed curves for the baryon component.
}
\label{fig:profile}
\end{figure}

In Fig.~\ref{fig:m-r relation} we show the mass-radius relation of DANSs with DM fraction $f_\chi =0.05$~\citep{2018PhRvD..97l3007E}. 
The calculations are done for two cases of DM interaction strengths ($y=0.1$ and $1000$) and two cases of DM particle mass ($m_\chi=0.1$ and $1\,{\rm GeV}$).
We see that in the case of small interaction strength ($y=0.1$), the DM effects on mass-radius are dominated by the DM particle mass, namely light DM particles ($m_\chi=0.1\,{\rm GeV}$) tend to increase the maximum mass. In contrast, heavy DM particles ($m_\chi=1\,{\rm GeV}$) would instead reduce the maximum mass of NSs.
In the case of large interaction strength ($y=1000$), regardless of the DM particle mass, one observes an increase in the total gravitation mass but nearly unchanged radii for the baryon components. 

The density profiles of $M_T=1.4\,\Msun$ DANSs are further reported in Fig.~\ref{fig:profile}, for both cases of DM interaction strengths and particle masses. The DM fraction is again fixed at $f_\chi=0.05$.The influences of the DM properties on the mass-radius relations of DANSs observed above in Fig.~\ref{fig:m-r relation} can then be understood as follows: 

1) In the case of small interaction strength ($y=0.1$), light DM particles ($m_\chi=0.1\,{\rm GeV}$) would form a dark halo, i.e., $R_D>R_B$. 
Therefore most DM gathers into the halo, thereby leaving only a small mass fraction inside the DANSs to impose a negligible influence on the stellar structure.
As a result, one observes an increase in the total gravitation mass, which comes from the halo mass, but nearly unchanged radii for the baryon components. 
The situation changes for heavy DM particles with $m_\chi=1\,{\rm GeV}$, as they form a dense core in the NS interior. 
Under this circumstance, the gravitational effect of the core cannot be neglected, and consequently, the radius of the baryon component shrinks.   

2) In the case of large interaction strength ($y=1000$), the repulsion could force DM to form a halo structure. In this case, we would again expect an increase in the total mass and an unchanged radius for the baryon component. As a result, the mass-radius relation could be indistinguishable for different DM particle masses. 
Nevertheless, it is possible to probe the $m_\chi$ value from the radius of DM component $R_D$, for example, through the measurements of tidal deformability $\Lambda$~\citep{2017PhRvL.119p1101A}, as it approximately scales with the fifth power of the stellar radius, i.e., $\Lambda\propto R_D^5$. 

\subsection{Dark core and dark halo}\label{subsec:halo/core}

As noted earlier, the distributions of DM in DANSs present two scenarios, namely dark halos ($R_D>R_B$) and dark cores ($R_D<R_B$). The formation of a dark core/halo depends not only on the DM properties but also on the DM amount.
The actual amount of DM accumulated in NSs should depend on the evolutionary history and the external environment of the NSs. See, e.g., \citet{2021PhRvD.103d3009K} for detailed discussions on the mixing of DM with ordinary matter in NSs. However, a thorough exploration of the accumulation mechanisms is still lacking and would be an interesting topic to explore in the future.
For the present purpose, we take two representative values of the DM fraction, $f_\chi=0.05$~\citep{2018PhRvD..97l3007E} and $f_\chi=10^{-4}$~\citep{2019JCAP...07..012N}.

\begin{figure}
\centering
\includegraphics[width=3.4in]{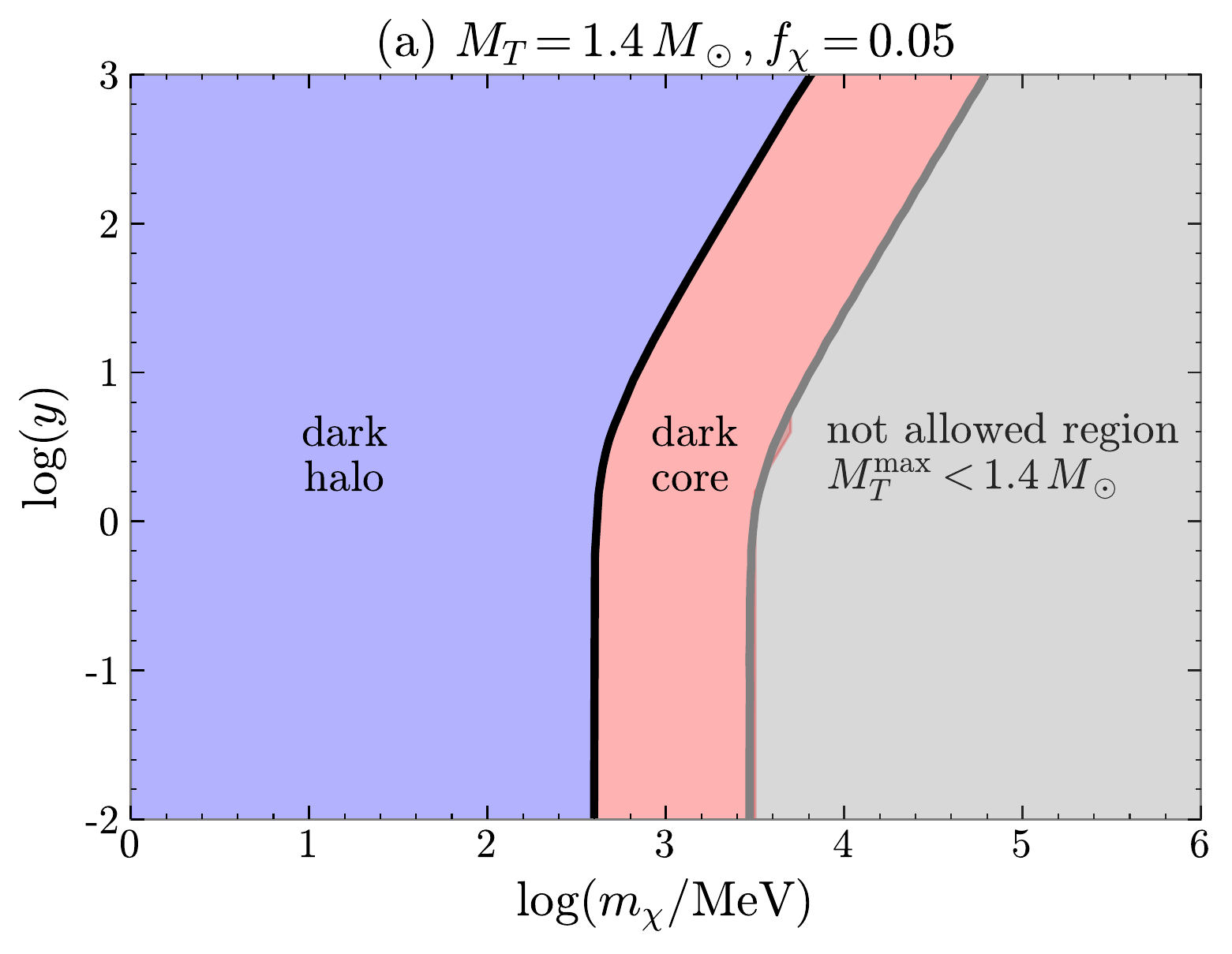}
\includegraphics[width=3.4in]{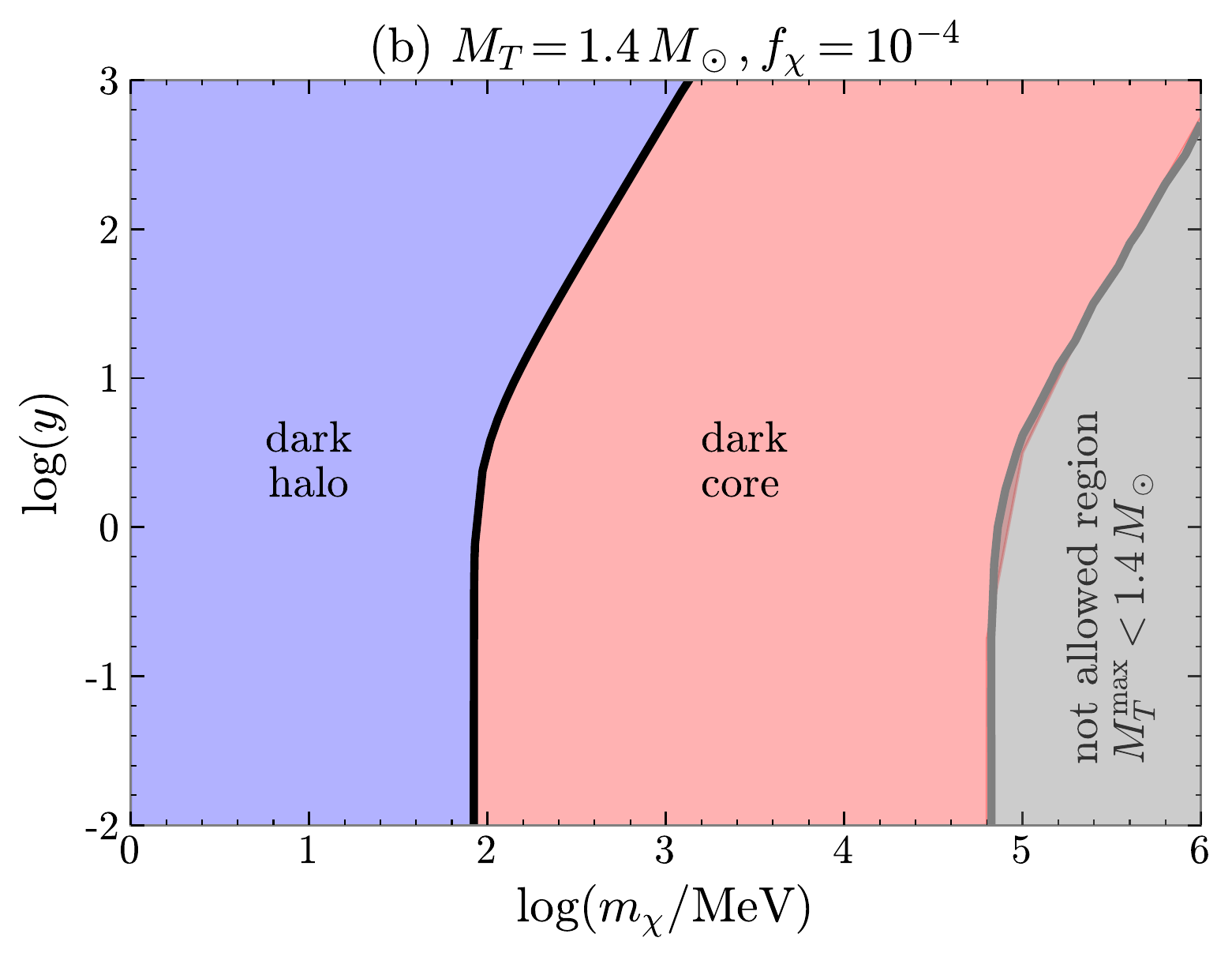}
\caption{DM parameter space of a dark core/halo for a $M_T=1.4\,M_\odot$ DANS with (a) $f_\chi=0.05$~\citep{2018PhRvD..97l3007E} (upper panel) and (b) $f_\chi=10^{-4}$~\citep{2019JCAP...07..012N} (lower panel). The grey shaded regions in two panels correspond to $M_T^{\rm max}<1.4\,M_\odot$ and thus are not allowed.}
\label{fig:param space}
\end{figure}

In Fig.~\ref{fig:param space} we show the $(m_{\chi},y)$ parameter spaces for DM particles forming a dark core/halo. The calculations are done with a fixed total gravitational mass $M_T=1.4\,M_\odot$.
As expected, for both cases of DM fractions, light/heavy DM particles tend to form a dark halo/core, and large/small interaction strengths tend to create a dark halo/core.
The solid black lines in Fig.~\ref{fig:param space} mark the borders of the purple dark-halo regions and the pink dark-core regions.
At small interaction strengths, the solid black lines are almost parallel to the $y$ axis, while at large interaction strengths, there is $m_\chi\propto y^{1/2}$.
When decreasing the DM fraction from $f_\chi=0.05$ (in the upper panel) to $10^{-4}$ (in the lower panel), the dark-halo (dark-core) region shrinks (expands), with the black border lines shifting slightly to the left.

\begin{figure*}
\centering
\includegraphics[width=3.4in]{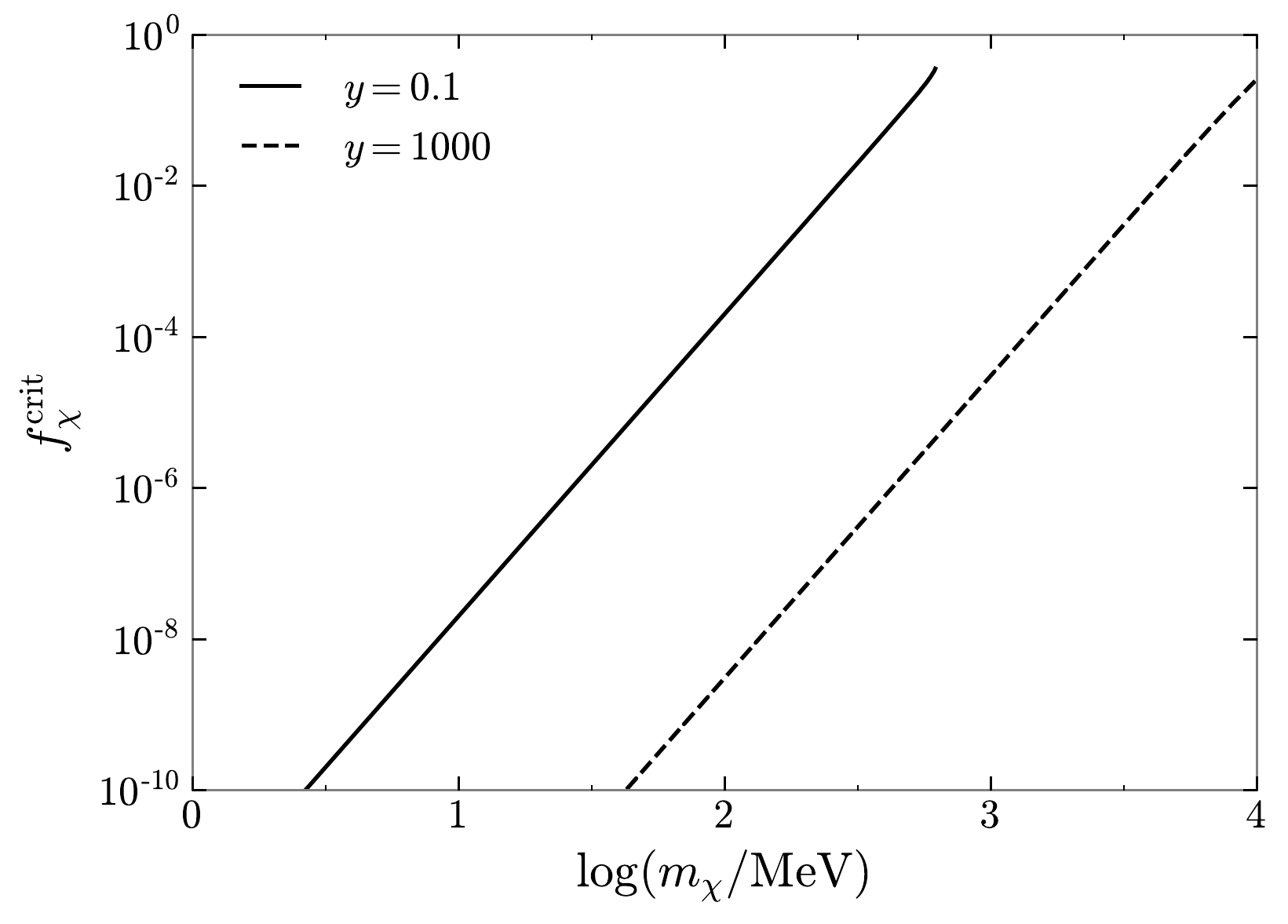}
\includegraphics[width=3.4in]{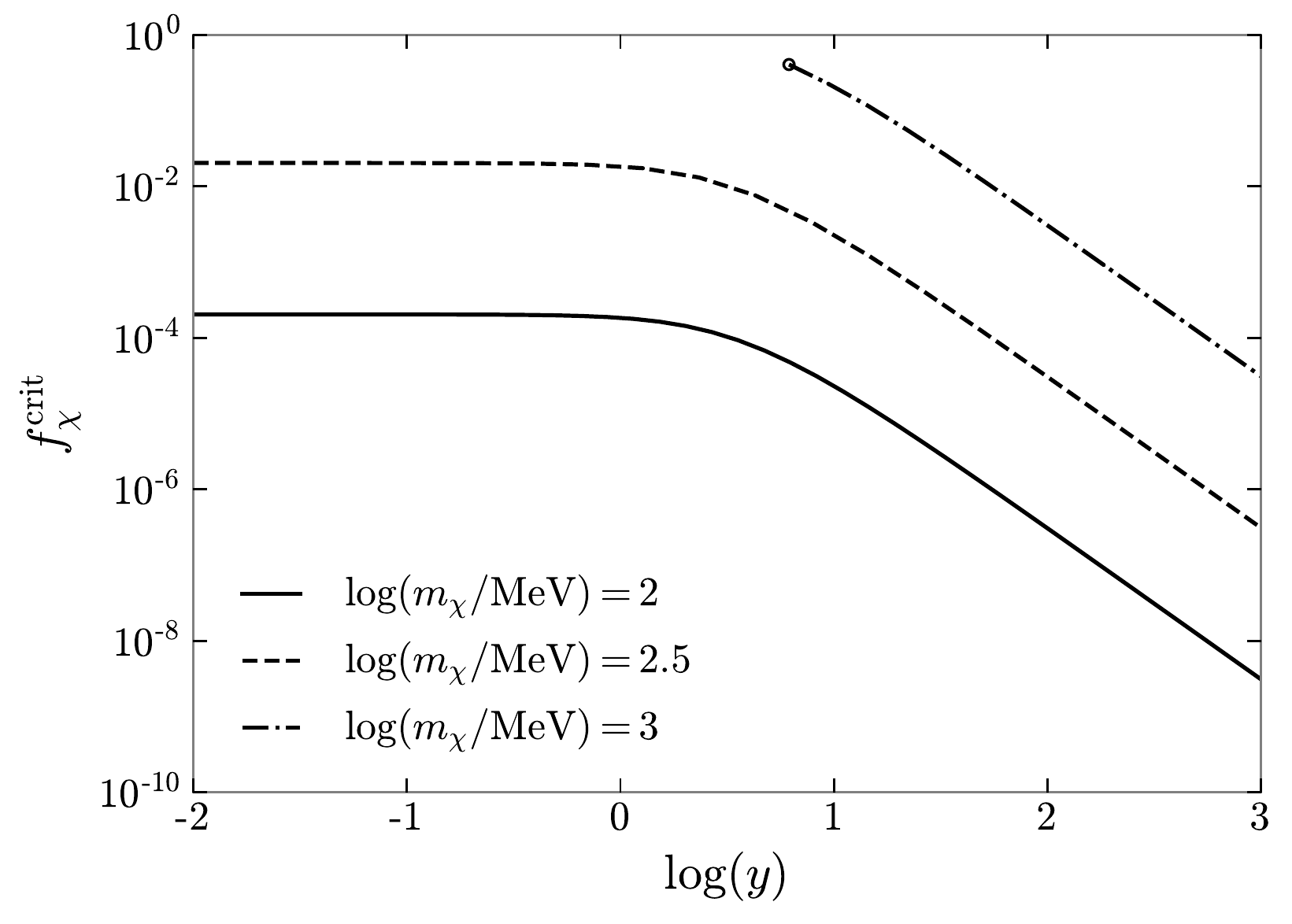}
\caption{(Left panel) Critical DM fraction as a function of DM particle mass for different interaction strengths; 
(Right panel) Critical DM fraction as a function of the interaction strengths for different DM particle masses. 
For each panel the DANS mass is fixed at $M_T=1.4\,\Msun$ and the baryon component is described by NL3$\omega\rho$ EOS. The dash-dotted line in the right panel ends at the circular symbol, due to small values of the interaction strength could not support a $M_T=1.4\,M_\odot$ DANS.
}\label{fig:fcrit}
\end{figure*}

For given DM properties, the critical DM fraction, $f_\chi^{\rm crit}$, can be calculated,
at which the radii of the baryonic matter and DM components coincide, i.e., $R_B=R_D$.
In Fig.~\ref{fig:fcrit}, we report the critical DM fractions as functions of the DM particle mass (left panel) and the interaction strength (right panel).
In the left panel, we see that there exists an excellent correlation between $m_\chi$ and $f_\chi^{\rm crit}$ as $f_\chi^{\rm crit}\propto m_\chi^4$. 
The right panel shows that $f_\chi^{\rm crit}\simeq{\rm const}$ for $y\ll1$ but $f_\chi^{\rm crit}\propto y^{-2}$ when $y\gg1$.
In the following, we further demonstrate and examine in detail the dependence of $f_\chi^{\rm crit}$ on the DM properties.

We first raise a new criterion for the formation of a dark core/halo. 
We begin with the chemical potential in the proper frame, $\mu_c$, which should be a constant throughout the equilibrium configuration~\citep{1972PThPh..47..444K}, i.e.,
\begin{equation}
    \mu_c = \mu \sqrt{g(r)} = {\rm const}\ ,
\end{equation}
where $\mu=\D \epsilon/\D n$ is the chemical potential in local Lorentz frame and $g(r)$ is the $(t,t)$ component of the metric. By defining the relativistic specific enthalpy $h$ as: 
\begin{equation}
  h=\int \frac{\D P}{\epsilon+P} = \int \frac{\D\mu}{\mu}\ ,  
\end{equation}
we find the derivative~\footnote{Another conclusion from this derivative is that the ratio of the two chemical potentials, $\mu_B/\mu_D$, is a constant throughout the stellar, see e.g.~\citet{2013PhLB..725..200G,2020PhRvD.102f3028I}.}: 
\begin{equation}
    \frac{\D h}{\D r} = -\frac{\D\ln\sqrt{g(r)}}{dr}\ ,
\end{equation}
is only a function of $r$. Therefore, the decreases of the enthalpies of two components, $h_B$ and $h_D$, must be equal in the same radial interval $[r,r+\D r]$. Combining with the fact that the decrease of $h_B/h_D$ will terminate once $h_B/h_D$ vanishes at the surface of baryonic matter/DM, we conclude that the formation of a dark core/halo is determined by the central enthalpies of the two components, i.e.,
\begin{equation}
\begin{split}
     h_D^{c}<h_B^c \quad \Rightarrow \quad{\rm dark\ core}\ ,\nonumber\\   
     h_D^{c}>h_B^c \quad \Rightarrow \quad{\rm dark\ halo}\ ,
\end{split}
\end{equation}
where $h_B^c$ and $h_D^c$ are the central enthalpies of the baryon and DM components, respectively.

With this criterion at hand, we proceed to derive the critical DM fraction of a DANS. 
In such critical cases, we denote the central enthalpy as $h_c\equiv h_B^c=h_D^c$ and the radius $R\equiv R_B=R_D$. 
For fermionic DM considered in the present paper, we have:
\begin{equation}\label{eq:hc-x}
    h_c = \ln\left(\sqrt{1+x^2}+\frac{2y^2}{3\pi^2}x^3\right)\ .
\end{equation}
Then the asymptotic solutions of $x$ can be found analytically for small and large interaction strengths, i.e.,
\begin{equation}\label{eq:x}
    x = 
  \left\{%
  \begin{array}{l}
  \sqrt{e^{2h_c}-1} \ , \qquad y\ll1\ , \\ 
  \left(\frac{3\pi^2e^{h_c}}{2y^2}\right)^{1/3}\ ,\quad y\gg1\ .
  \end{array}
  \right.
\end{equation}

For current observed NSs with $M\sim1-2\,M_\odot$, the central entralpy is $h_c\sim0.1-0.3$.
Under such a range of $h_c$, the DM energy density $\epsilon_\chi$ is dominated by the rest-mass term (see Appendix~\ref{sec:Appendix A} for detailed discussions), i.e., $\epsilon_\chi\simeq m_\chi n_\chi=m_\chi^4 x^3/(3\pi^2)$. Then the amount of DM could be estimated as $M_D\simeq \epsilon_\chi R^3=x^3m_\chi^4R^3/(3\pi^2)$, or equivalently, 
\begin{equation}\label{eq:critical fraction}
    f_\chi^{\rm crit}  \simeq 5\times10^{-7}\left(\frac{x}{0.1}\right)^3\left(\frac{m_\chi}{100{\rm MeV}}\right)^4\left(\frac{1.4\Msun}{M_T}\right)\left(\frac{R}{12{\rm km}}\right)^3\ .
\end{equation}

We here mention that Eq.(\ref{eq:critical fraction}) together with Eq.(\ref{eq:x}) can be used to explain all the features previously observed for $f_\chi^{\rm crit}$ in Fig.~\ref{fig:param space} and Fig.~\ref{fig:fcrit}:
\begin{itemize}
\item 
From Eq.(\ref{eq:critical fraction}), $m_\chi\propto x^{-3/4}$ at fixed $f_\chi^{\rm crit}$; Then from Eq.(\ref{eq:x}), $m_\chi$ does not rely on the value of $y$ if $y\ll1$, whereas $m_\chi\propto y^{1/2}$ if $y\gg1$, as shown in Fig.~\ref{fig:param space};
\item 
Obviously from Eq.(\ref{eq:critical fraction}), $f_\chi\propto m_\chi^4$, as shown in Fig.~\ref{fig:fcrit} (in the left panel); 
\item 
Substituting $x$ in Eq.(\ref{eq:critical fraction}) by Eq.(\ref{eq:x}), one can easily find that $f_\chi$ is independent of $y$ for $y\ll1$, while $f_\chi\propto y^{-2}$ for $y\gg1$, as shown in Fig.~\ref{fig:fcrit} (in the right panel). 
\end{itemize}

\section{Dark-halo effects on the pulsar pulse profile}\label{sec:pulse profile}

\begin{table*}
	\centering
	\caption{Properties of two constructed DANS models, by fixing $M_T(R_B)\equiv M_B(R_B)+M_D(R_B)=1.4\,M_\odot$ and $M_{\rm halo}\equiv M_T-M_T(R_B)=0.15\,M_\odot$ but varying the DM particle mass $m_\chi$ and the interaction strength $y$. The corresponding pure NS cases, which share the same $M_T(R_B)$ and $R_B$ with the two DANSs, respectively, are also listed.}
	\label{tab:DANS models}
\renewcommand\arraystretch{1.2}
  \begin{ruledtabular}
\begin{tabular}{c|cccc|ccccc} 
Model    & $m_\chi/{\rm GeV}$  &$\log(y)$ &$M_T(R_B)/M_\odot$  &$M_{\rm halo}/M_\odot$ &$M_T/M_\odot$  &$f_\chi$ & $R_B/{\rm km}$ &$R_D/{\rm km}$ &$M_{\rm halo}/R_D$ \\
\hline 
DANS I    &0.1 &1.5  &1.4 &0.15 &1.55 &0.096  &13.75 &2771.4   &$8.0\times10^{-5}$      \\
pure NS I    &-- &--  &1.4 &0 &1.4 &0 &13.75 &--   &0      \\
\hline
DANS II  &1   &1.5  &1.4 &0.15 &1.55 &0.173 &13.23 &28.7  & $7.7\times10^{-3}$   \\
pure NS II  &-- &--  &1.4 &0 &1.4 &0 &13.23 &--   &0      \\
\end{tabular}
  \end{ruledtabular}
\end{table*}
Based on the oblate Schwarzschild approximation for the NS spacetime~\citep{2007ApJ...663.1244M}, the pulse-profile modeling technique has been used by the Neutron Star Interior Composition Explorer (NICER)~[see \citet{2016RvMP...88b1001W,2019AIPC.2127b0008W} and references therein] to deliver tight constraints on the mass and radius of NSs~\citep{2019ApJ...887L..21R,2021ApJ...918L..27R,2019ApJ...887L..24M,2021ApJ...918L..28M} and to pin down the unknown dense matter 
EOS~\citep{2019ApJ...887L..24M,2019ApJ...887L..22R,2021ApJ...918L..29R}.
Since the pulse profile may be modified due to the extra gravitational potential of the dark halo~\citep{2022PhRvD.105f3005D}, 
in this section, we numerically calculate the pulse profiles of DANSs and investigate the effects resulting from the existence of a dark halo.

In this first attempt at studying the dark-halo effects on the pulse profiles, several simplifications are in order.
We assume the radiation emitted from two point-like spots sitting oppositely on the baryon surface (i.e., at $r=R_B$). And the specific intensity of the radiation is assumed to be isotropic~\citep{1998ApJ...499L..37M}. 
We do include the Doppler boost and relativistic aberration in our calculation but neglect the frame-dragging and the stellar deformation due to rotation.
We also assume that the distance between the DANS and the observer is large enough to be treated mathematically as infinity. Below we introduce the necessary formalism for the calculations.

The observed differential flux at distance $D$ from a point-like spot is given by~\citep[see e.g.,][for a detailed derivation]{2006MNRAS.373..836P}:
\begin{equation}
    \D F = \delta^5g(R_B)I^\prime(\nu^\prime)\D \nu^\prime\cos\alpha\frac{\D\cos\alpha}{\D\cos\psi}\frac{\D S^\prime}{D^2}\ ,
\end{equation}
where $\delta$ is the Doppler factor. $I^\prime(\nu^\prime)$ is the radiation intensity at the comoving frame. 
$\alpha$ is the emission angle and $\psi$ is the angle between the local radial direction and the line of sight. $dS^\prime$ is the proper differential area where the photons are emitted (see Fig.~1 in~\citet{2006MNRAS.373..836P} for illustration).

For further analysis, we use the bolometric flux normalized by $\int I^\prime(\nu^\prime)\D \nu^\prime\D S^\prime/D^2$,
\begin{equation}\label{eq:flux}
    F \equiv \frac{D^2\int \D F}{\D S^\prime \int I^\prime(\nu^\prime)\D\nu^\prime} = \delta^5g(R_B)\cos\alpha\frac{\D\cos\alpha}{\D\cos\psi}\ .
\end{equation}
Here Eq.(\ref{eq:flux}) accounts for the special realtivistic effects (Doppler boost and relativistic aberration), the gravitational redshift and light bending.

To obtain the normalized flux, one need to determine the value of $\psi$ and the relation between $\psi$ and $\alpha$. The later can be derived from the photon geodesic equation as (see Appendix~\ref{sec:Appendix B} for derivation):
\begin{equation}\label{eq: psi-alpha}
    \psi = \int_{R_B}^\infty\frac{\sqrt{fg}}{r^2}\left[\frac{1}{b^2}-\frac{g}{r^2}\right]^{-1/2}\D r\ ,
\end{equation}
where $b=R_B\sin\alpha/\sqrt{g(R_B)}$ is the impact parameter. $f(r)$ is the $(r,r)$ component of the metric.

On the other hand, $\psi$ can be obtained from the geometry relation:
\begin{equation}
    \cos\psi = \cos i\cos \theta + \sin i\sin \theta \cos(2\pi\nu t)\ ,
\end{equation}
where $i$ is the inclination angle of the spin axis to the line of sight, $\theta$ is the spot colatitude and $\nu$ is the rotation frequency. Here $t = 0$ is chosen when the spot is closest to the observer. 

With $\psi$ at hand, we invert Eq.~(\ref{eq: psi-alpha}) to get $\alpha$. Then we can calculate the Doppler factor $\delta$ and $\D(\cos\alpha)/\D(\cos\psi)$ and obtain the normalized flux. Finally, the time-delay effects caused by different traveling paths of the emitted photons should be considered (see Appendix~\ref{sec:Appendix B} for derivation):
\begin{equation}
    \Delta t = \int_{R_B}^\infty\sqrt{\frac{f}{g}}\left[\left(1-\frac{b^2g}{r^2}\right)^{-1}-1\right]\D r\ .
\end{equation}
We correct the observer time as $t_{\rm obs} = t+\Delta t$.

For the following study, we define two new masses, namely, (i) $M_T(R_B)\equiv M_B(R_B)+M_D(R_B)$, which corresponds to the total gravitational mass inside the baryon radius $R_B$; (ii) $M_{\rm halo} \equiv M_T-M_T(R_B)$, which corresponds to the mass of the dark halo.
It is obvious that $M_T(R_B)=M_T$ in the absence of a dark halo; otherwise, $M_T(R_B)<M_T$. Notice that the original Schwarzschild spacetime outside a pure NS with no halo would be altered when adding a dark halo. 
The change can be characterized by the increase of $g(R_D)$ as:
\begin{equation}
    \Delta g(R_D) = \left(1-\frac{2M_T(R_B)}{R_D}\right)-\left(1-\frac{2M_T}{R_D}\right)=\frac{2M_{\rm halo}}{R_D}\ .
\end{equation}
Hereafter we choose $M_{\rm halo}/R_D$ as a characteristic parameter and will show that the modification of the pulse profile depends sensitively on $M_{\rm halo}/R_D$.

\begin{figure}
\centering
\includegraphics[width=3.4in]{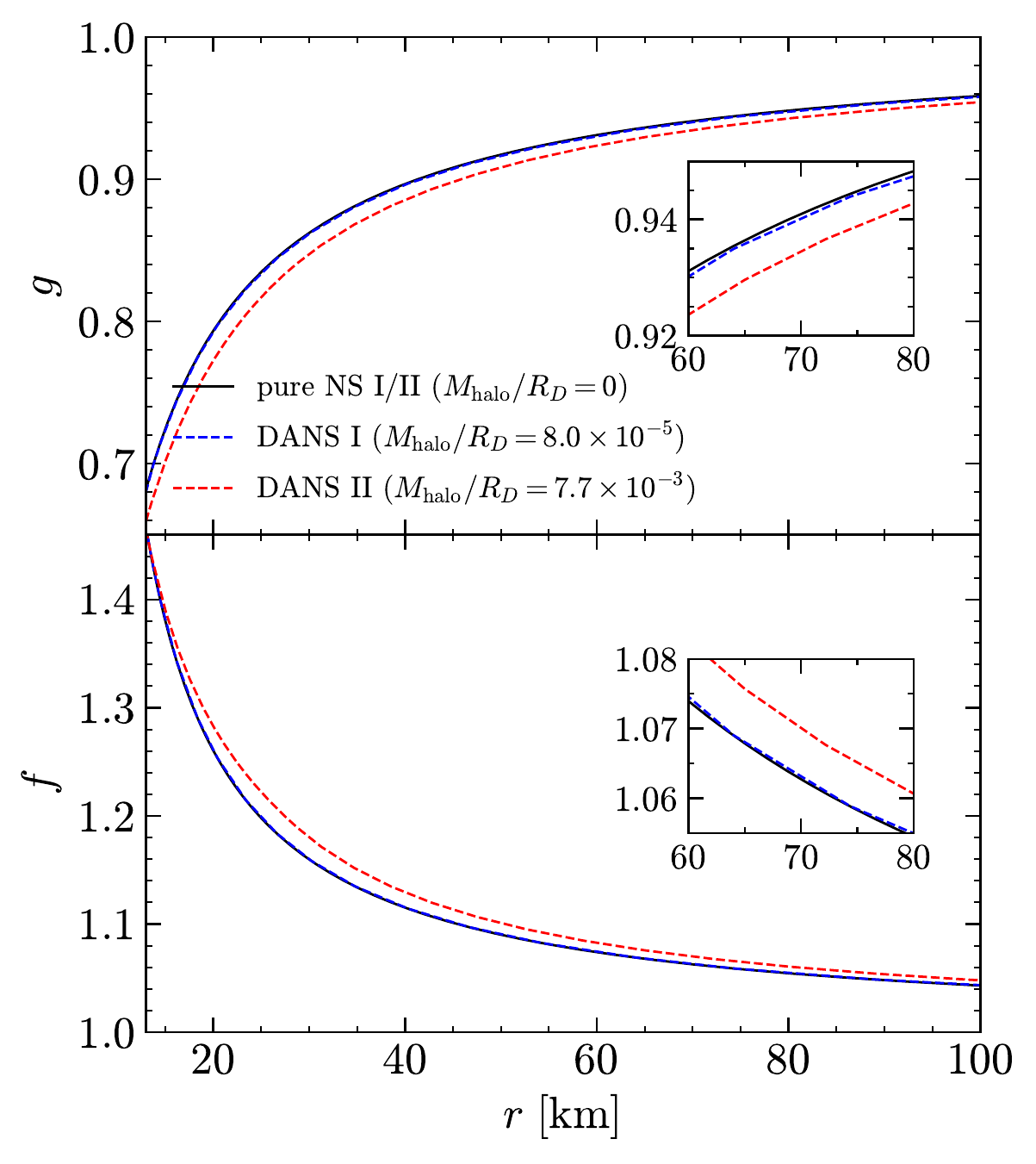}
\caption{Metric components $g(r)$ and $f(r)$ as functions of the radius $r$ for DANS models listed in Table~\ref{tab:DANS models}. 
The Schwarzschild metrics outside a pure NS with $M_T=1.4\,M_\odot$, $g=1-2M_T/r$ and $f=(1-2M_T/r)^{-1}$, is also shown with black solid curves.
} 
\label{fig:metric}
\end{figure}

\begin{figure}
\centering
\includegraphics[width=3.4in]{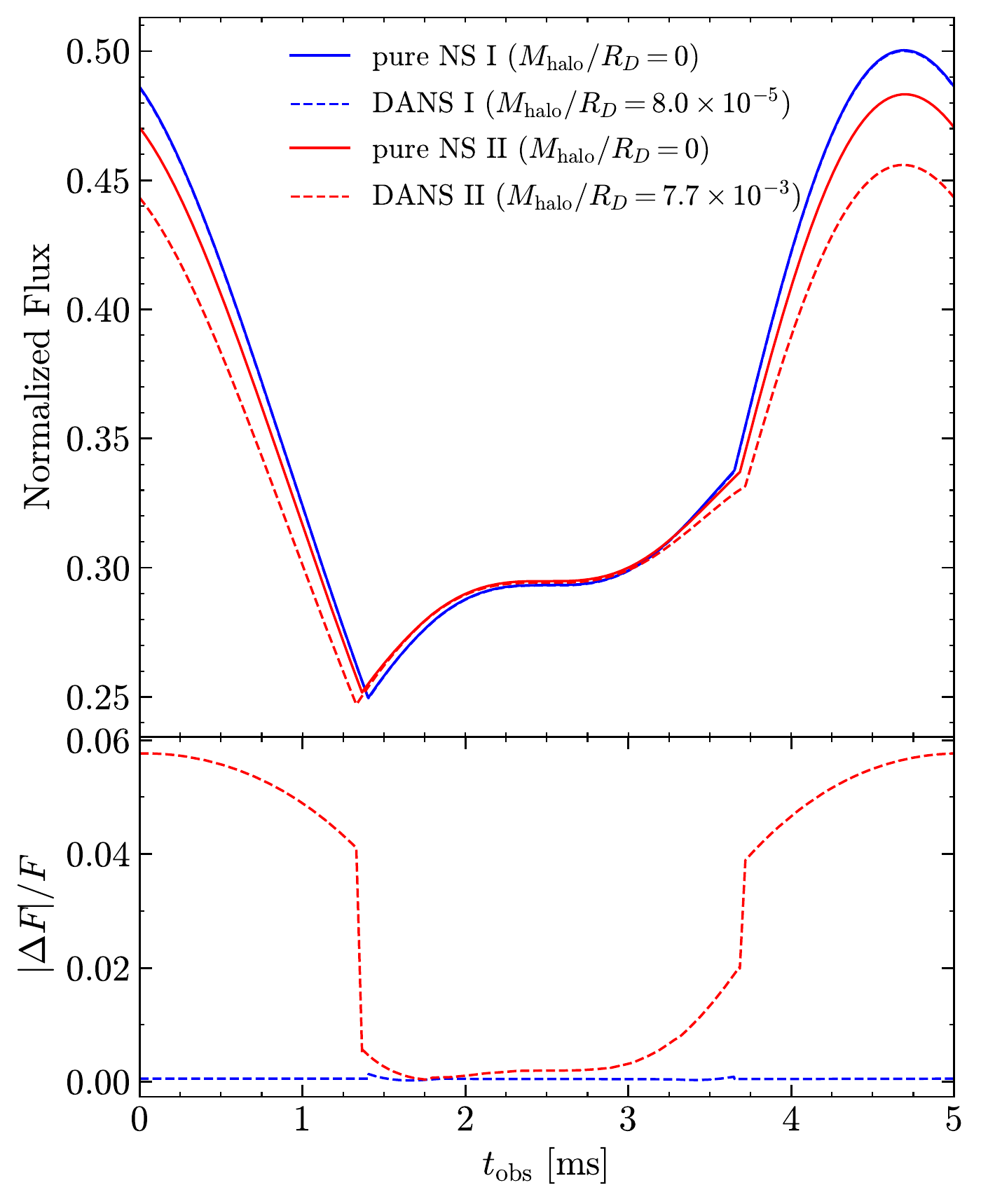}
\caption{Pulse profiles from two point-like spots for the DANSs listed in Table~\ref{tab:DANS models} (dashed curves). The spot colatitude at northern hemisphere is $\theta =\pi/4$ and the inclination angle of the spin axis to the light of sight $i=\pi/4$. The rotating frequency is taken to be $\nu =200\,{\rm Hz}$.
The corresponding pure NS cases with same $M_T(R_B)$ and $R_B$ are also shown (solid curves). 
Note the two pure NS cases are not the same, due to the small difference in the radii of the baryon component $R_B$ (see Table~\ref{tab:DANS models}).
}
\label{fig:pulse profile}
\end{figure}

In Table~\ref{tab:DANS models} we construct two DANSs by fixing $M_T(R_B)=1.4\Msun$ and $M_{\rm halo}=0.15\,M_\odot$ but varying the DM particle properties. Their corresponding pure NSs, which share the same $M_T(R_B)$ and $R_B$ with them, respectively, are also listed. It should be emphasized that the values of $M_{\rm halo}/R_D$ are quite different for the two constructed DANSs.
In Fig.~\ref{fig:metric}, we plot the metric as a function of the radius. We note that the metric functions of the DANSs deviate from their corresponding pure NS ones. Meanwhile, the deviation close to the star can be characterized by $M_{\rm halo}/R_D$ since they are of the same order of magnitude. 

In Fig.~\ref{fig:pulse profile} we show the pulse profiles for the DANSs and pure NSs listed in Table~\ref{tab:DANS models}. We take the spot colatitude $\theta=\pi/4$, the inclination angle $i=\pi/4$ and the rotating frequency $\nu=200\ {\rm Hz}$ in calculation.
It is noticeable that the pulse profiles of DANSs deviate from the pure NS cases. The difference appears mainly in the amplitudes of the normalized flux, whereas the phase shift is almost negligible. The peak flux deviation is about 6\% for model DANS II ($M_{\rm halo}/R_D=7.7\times10^{-3}$) but only 0.05\% for model DANS I ($M_{\rm halo}/R_D=8\times10^{-5}$). The nearly two orders of magnitude difference between the peak flux deviations is close to the difference between $M_{\rm halo}/R_D$, implying that the peak flux deviation may strongly depend on $M_{\rm halo}/R_D$.

\begin{figure}
\centering
\includegraphics[width=3.4in]{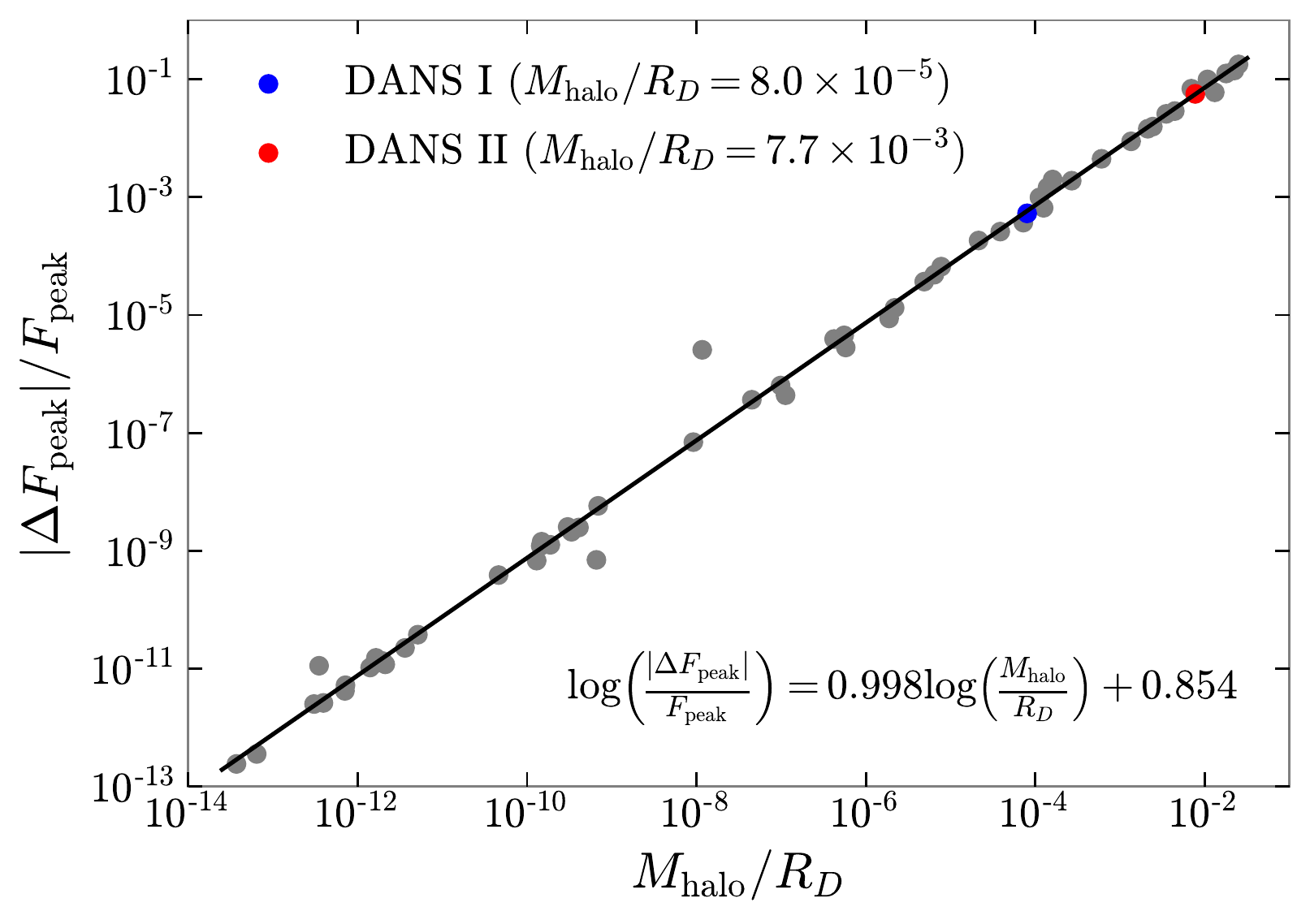}
\caption{Peak flux deviation as a function of $M_{\rm halo}/R_D$. The grey dots correspond to the results from dozens of DANS models, which are generated with parameters chosen randomly throughout a wide parameter space (see text for details). The blue and red dots are results from the two DANSs shown in Fig.~\ref{fig:pulse profile}, respectively. The empirical relation is shown with a solid black line.
} 
\label{fig:Dg-DF}
\end{figure}

To clarify this dependence, in Fig.~\ref{fig:Dg-DF}, we show in detail the peak flux deviation as a function of $M_{\rm halo}/R_D$ for dozens of DANS models. These DANSs are generated with parameters chosen randomly throughout the parameter spaces of $\log(m_\chi/{\rm MeV})\in[0,6]$, $\log(y)\in[-2,3]$, $M_T\in[1,2.5]\,M_\odot$, $f_\chi\in[0,0.5]$, $i\in[\pi/6,\pi/2]$, $\theta\in[\pi/6,\pi/2]$ and $\nu\in[100,400]\,{\rm Hz}$. We observe an approximate linear correlation between 
the peak flux deviation and $M_{\rm halo}/R_D$ in logarithmic space, i.e.,
\begin{equation}
    \log\left(\frac{|\Delta F_{\rm peak}|}{F_{\rm peak}}\right)=0.998\log\left(\frac{M_{\rm halo}}{R_D}\right)+0.854\ .
\end{equation}
with the corresponding coefficient of determination $R^2=0.995$.
It is seen that for compact dark halos with $M_{\rm halo}/R_D\simeq 0.01$, the deviations of peak flux can reach about 10\%.
Therefore it may be relevant for the NICER analysis of the mass and radius of X-ray pulsars for the study of dense matter EOS, and we also expect that such effects can be detected with future X-ray data.
We mention here that, although our study provides a connection of the observed flux with the microscopic physics of DM and baryonic matter, it is only a first, crude approximation to the phenomena occurring in the observed flux of X-ray pulsars, see discussions in, e.g.,~\citet{2013ApJ...776...19L,2014ApJ...787..136P,2015ApJ...808...31M}. 
A more detailed analysis incorporating, e.g., the number of photon counts and the background emission, will be performed in the future for the observability related to DM halo.

\section{Dark matter particle properties}\label{sec:dark matter properties}

In this section, we perform Bayesian parameter estimation on the unknown DM parameters, using the available NICER mass-radius measurements of two X-ray pulsars, i.e., PSR J0030+0451~\citep{2019ApJ...887L..24M,2019ApJ...887L..21R} and PSR J0740+6620~\citep{2021ApJ...918L..28M,2021ApJ...918L..27R}. 
As illustrated above (see Sec.~\ref{subsec:halo/core}), $h_c$ plays an important role in determining the core/halo configuration. 
Therefore, it is convenient to combine the central enthalpy of baryonic matter $h_B^c$ and the ratio of the two enthalpies $q\equiv h_D^c/h_B^c$, as well as the DM parameters $m_\chi$ and $y$, into a vector $\boldsymbol\theta$.
The total likelihood function in our analysis is expressed as: 
\begin{equation}
    p\left(\{M_i,R_i\}|\boldsymbol\theta\right) = \prod_i p_i\left(M_{T,i}(R_B),R_{B,i}|\boldsymbol\theta\right)\ ,
\end{equation}
where we equate the individual likelihood $p_i$ of observation $i$ to the joint posterior distributions of $M_i$ and $R_i$ reported by NICER~\citep{2019ApJ...887L..21R,2021ApJ...918L..27R}. 
Since the pulse-profile modeling of NICER should be modified for DANSs and one could not apply the NICER measurements to study DANSs directly.
We simply treat $M_{T,i}(R_B)$ and $R_{B,i}$ as the measured mass and radius.
Such treatment should be valid as long as $M_{\rm halo}/R_D$ is sufficiently small, and we consider two cases being consistent with it, i.e., (i) $M_{\rm halo}/R_D\leq10^{-3}$; thus, the corresponding flux deviations are smaller than $\sim1\%$, and (ii) $M_{\rm halo}/R_D=0$, which is equivalent to a dark core scenario.

We take logarithmic uniform distributions for the DM particle mass ($\log(m_\chi/{\rm MeV})\sim U[0,6]$) and for the interaction strength ($\log(y)\sim U[-2,3]$). 
The priors of $h^c_B$ are uniform distributed in the range of $[0.1,0.5]$ and $[0.2,0.6]$ for PSR J0030+0451 and PSR J0740+6620, respectively. We also set  uniform distributions for both cases of $M_{\rm halo}/R_D\leq10^{-3}$ ($q\sim U[0,3]$) and for the case of $M_{\rm halo}/R_D=0$ ($q\sim U[0,1]$). Note that $q\leq1$ corresponds to a dark core.

\begin{figure}
\centering
\includegraphics[width=3.4in]{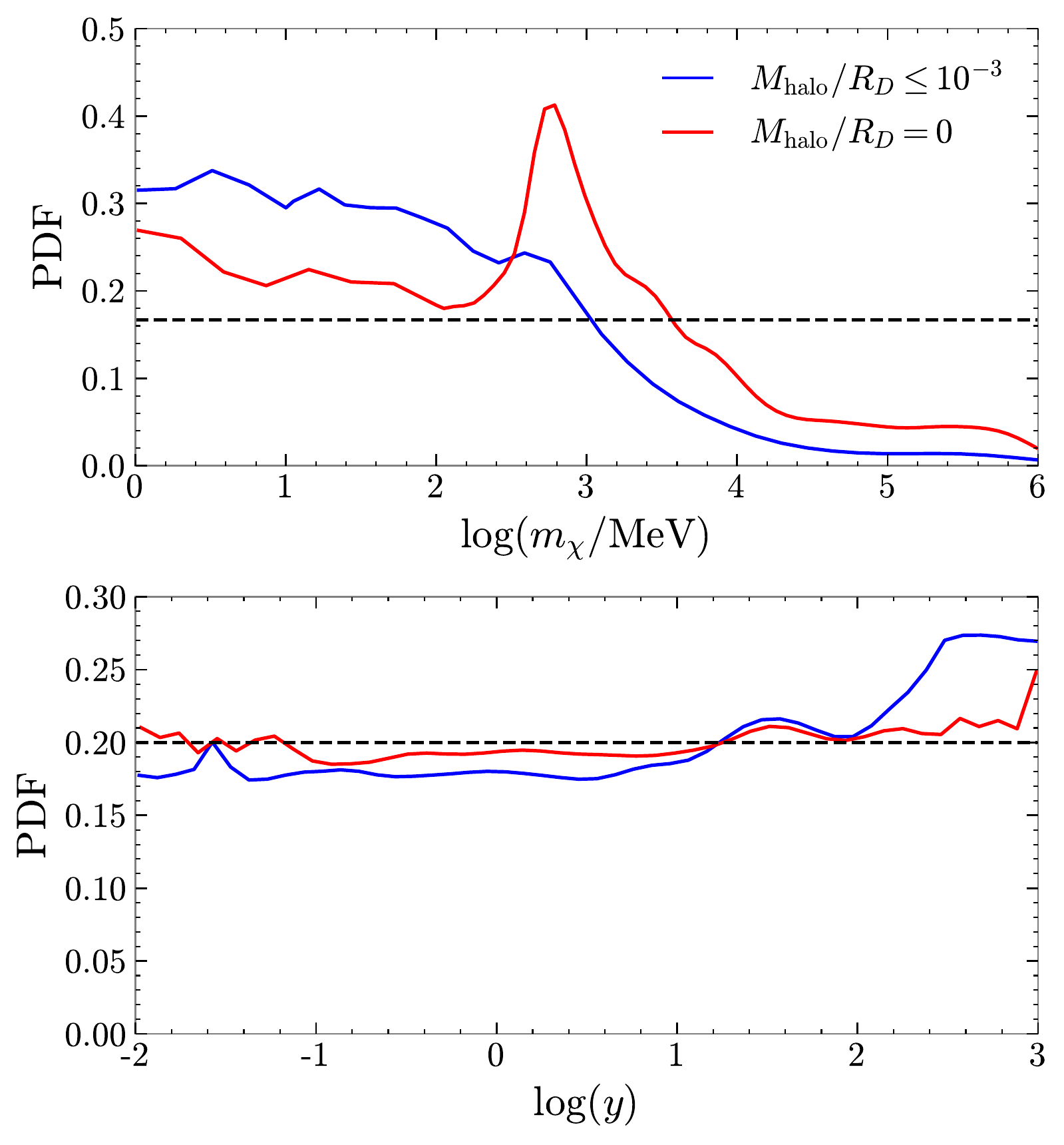}
\caption{Posterior distributions of the DM particle mass (upper panel) and the interaction strength (lower panel). The corresponding priors are shown as black dashed lines.
} \label{fig:PDF}
\end{figure}

With the priors and the likelihood function at hand, we sample from the posterior density using the nested sampling software \textsc{PyMultiNest}~\citep{2014A&A...564A.125B}. The resulting probability density functions (PDFs) of the DM particle mass and of the interaction strength are shown in Fig.~\ref{fig:PDF}. 

The obtained PDFs of the DM particle mass are quite different for the two cases. 
Assuming $M_{\rm halo}/R_D\leq10^{-3}$, the observational data prefers light DM particles and we can place a 90\% upper limit of $m_\chi\leq1.5\ {\rm GeV}$.  
Heavier particles are mostly excluded because they tend to reduce the maximum mass of DANS below the masses of PSR J0030+0451 or PSR J0740+6620. 
Whereas assuming $M_{\rm halo}/R_D=0$, there shows a prominent peak at $m_\chi\sim0.6\,{\rm GeV}$. The peak can be understood by considering the requirement of a dark core: 
Light DM particles more easily form a dark halo and thus are more likely to be rejected in the nested sampling process.
At the same time, to support the masses of observed pulsars, $m_\chi$ favors those values which are not too large. 
Nevertheless, the interaction strength $y$ is poorly constrained by the observations of the pulsar pulse profile; due to that, within the current modeling of DM particles, the DM rest-mass term dominates the DM energy and pressure (see Appendix.~\ref{sec:Appendix A}). 
Nevertheless, as mentioned above, since the interaction strength could affect the tidal deformability via affecting the radius of the DM component, future tidal deformability measurements from gravitational waves hold the promise of providing a complementary constraint on the interaction strength.  

\section{Summary and Conclusions}\label{sec:summary}

In this paper, we study how DM, admixed with ordinary baryonic matter in different distributions, changes the structure and observations of NSs and confront the DANS properties with NS observational data.
We consider DM as MeV-GeV self-interaction fermions and explore in detail the two possible scenarios, namely the dark halo and the dark core.  

Although the formation of a dark core/halo depends complicatedly on the DM particle mass ($m_{\chi}$) and self-interacting strength ($y$) as well as the fraction of DM accumulated ($f_{\chi}$). 
We here show that these dependencies can be understood in a unified manner from a newly-proposed criterion for identifying the formation of a dark core or a dark halo. That is, the formation of a dark core/halo is essentially determined by the central enthalpies of the baryonic and dark components. 
We further provide a general formula for the critical amount of DM, $f_\chi^{\rm crit}$, which is well consistent with the numerical results of solving the two-fluid TOV equations. 
In particular, $f_\chi^{\rm crit}$ correlates linearly with the fourth power of the DM particle mass, i.e., $f_\chi^{\rm crit}\propto m_\chi^4$;
Also, $f_\chi^{\rm crit}$ is insensitive to $y$ in the weak interacting case ($y\ll1$), but proportional to $y^{-2}$  in the strong interacting case ($y\gg1$).

For the first time, we also estimate the dark-halo effects on NS pulse profiles, which are observed in NICER-like X-ray missions.
Our analysis reveals that the effects depend sensitively on the value of $M_{\rm halo}/R_D$, 
and the modification in the peak flux deviation can reach up to $\sim$10\% if the dark halo is dense enough with $M_{\rm halo}/R_D\simeq 0.01$. 
Furthermore, we show that the existence of a dark halo could increase the maximum mass while a dense dark core could dramatically decrease it, which is in agreement with previous works. 

In addition, we apply the available NICER data to the DANS study and present a Bayesian parameter estimation of DM properties.
In the dark-core scenario ({$M_{\rm halo}/R_D=0$}), we find a peak PDF for the DM particle mass at $m_\chi\simeq0.6\,{\rm GeV}$;
While in the dark-halo scenario, there is a more loose upper limit of $m_\chi\leq1.5\,{\rm GeV}$ at 90\% confidence level when assuming $M_{\rm halo}/R_D\leq10^{-3}$.
No new constraint is found for the interaction strength between DM particles by analyzing the present pulse observations of PSR J0030+0451 and PSR J0740+6620 from NICER.

\appendix
\section{The dimensionless Fermi momentum and energy density of dark matter}\label{sec:Appendix A}

\begin{figure}
\centering
\includegraphics[width=3.4in]{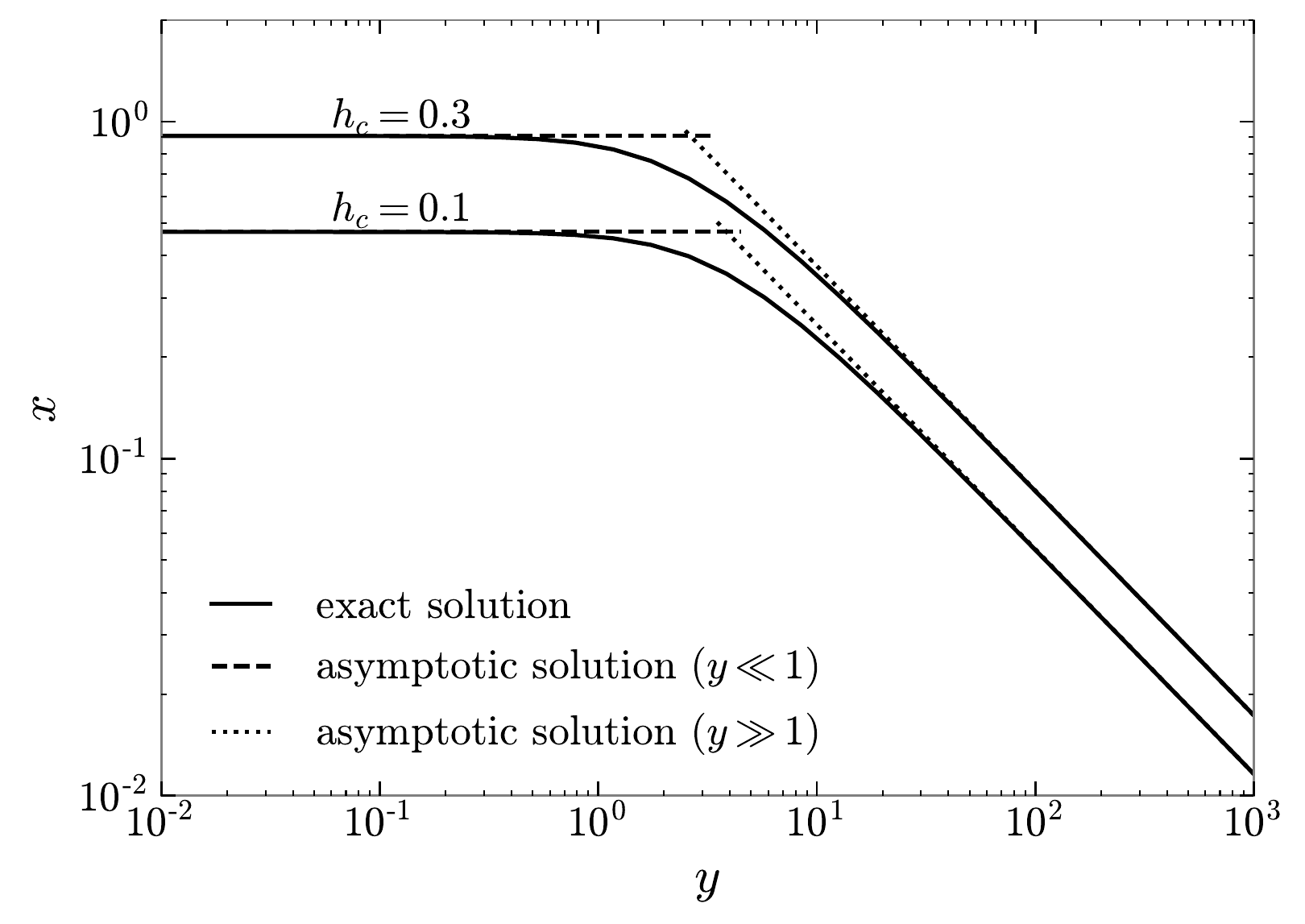}
\caption{Dimensionless Fermi momentum $x$ as a function of the interaction strength $y$ for $h_c=0.1$ and $h_c=0.3$. Both the exact solutions of Eq.~(\ref{eq:hc-x}) and the asymptotic solutions of Eq.(\ref{eq:x}) are shown.
} \label{fig:solution of x}
\end{figure}

In Fig.~\ref{fig:solution of x}, we report the exact solutions of the dimensionless Fermi momentum $x$ from Eq.~(\ref{eq:hc-x}), together with the asymptotic solutions of Eq.~(\ref{eq:x}). 
It is shown that the asymptotic solutions are sufficiently accurate for $y\lesssim0.1$ or $y\gtrsim10$. 
For the considered range of $h_c\simeq0.1-0.3$ relevant to the present study,
$x$ always lies in the range from $\mathcal{O}(0.01)$ to $\mathcal{O}(0.1)$.

\begin{figure}
\centering
\includegraphics[width=3.4in]{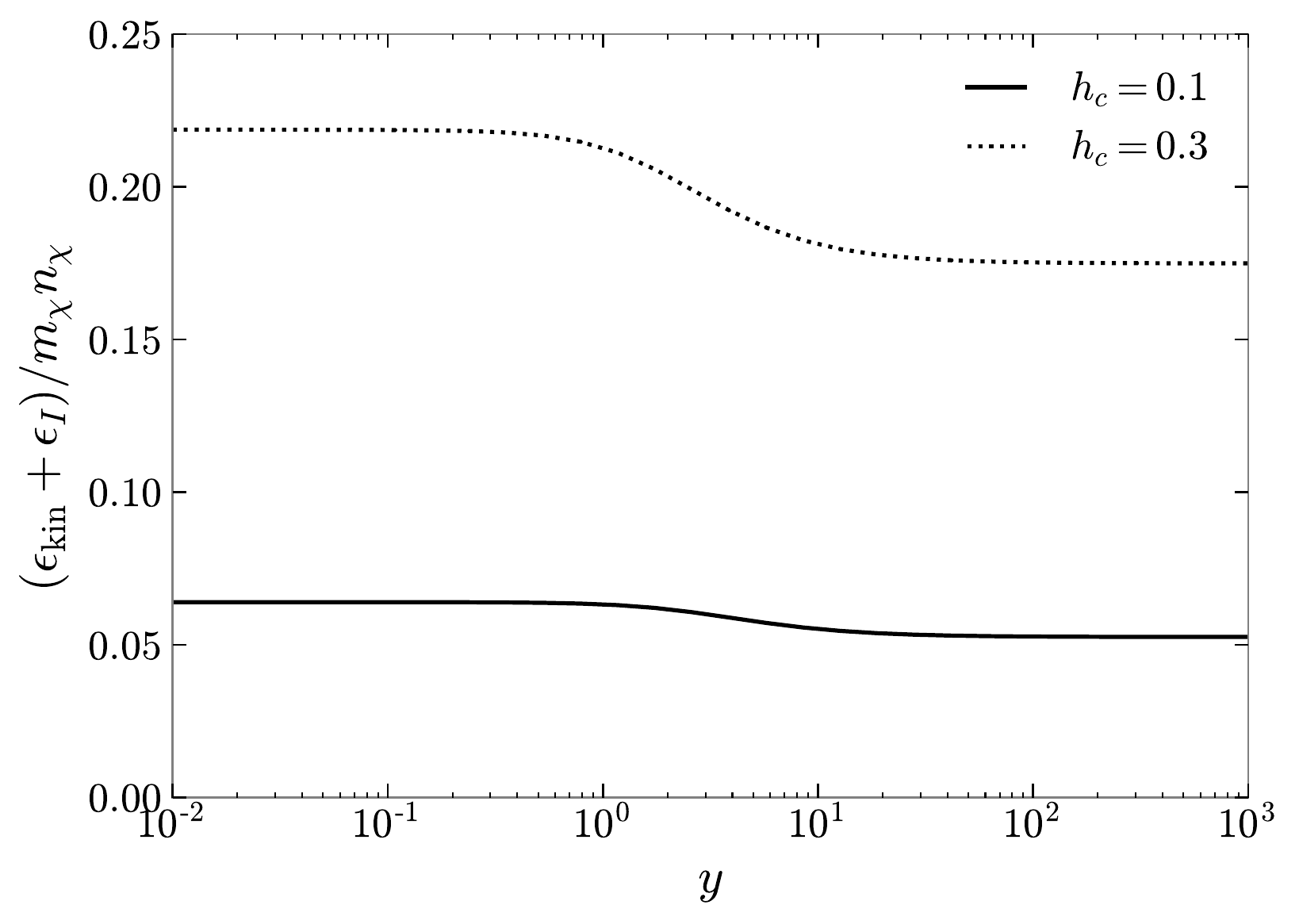}
\caption{Kinetic and interaction energy density (scaled by the rest mass density) as a function of the interaction strength for $h_c=0.1$ and $h_c=0.3$. 
} \label{fig:edensity comparison}
\end{figure}

Substituting $x$ into Eq.(\ref{eq:e_chi}), we can calculate the energy density of DM. In Fig.~\ref{fig:edensity comparison} we plot the kinetic and interaction energy density as a function of the interaction strength. We find that in the considered range of $h_c$, the kinetic and interaction energy density would not exceed $\sim20\%$ of the rest-mass density. 
This justifies the approximation $\epsilon_\chi\simeq m_\chi n_\chi$ we use in the context.


\section{Light bending and time delay}\label{sec:Appendix B}

Since we neglect the rotation effect on the spacetime, the metric is spherically symmetric and can be written as 
\begin{equation}
    \D s^2=-g(r)\D t^2+f(r)\D r^2+r^2\D \theta^2+r^2\sin^2\theta\D \phi^2\ .
\end{equation}
The spherical symmetry allows us to confine the photon geodesic to the equatorial plane (i.e., $\theta=\pi/2$). Then one can choose an affine parameter $\lambda$ such that
\begin{align}
    &\frac{\D t}{\D\lambda} = \frac{1}{g}\ ,\label{eq:dtdl}\\ 
    &\frac{\D r}{\D\lambda} = f^{-1/2}\left(\frac{1}{g}-\frac{b^2}{r^2}\right)^{1/2}\ ,\label{eq:drdl}\\
    &\frac{\D\theta}{\D\lambda} = 0\ ,\\
    &\frac{\D\phi}{\D\lambda} = \frac{b}{r^2}\ ,\label{eq:dphidl}
\end{align}
where $b$ is the impact parameter. The impact parameter is related to the emission angle at the baryon surface as $b=R_B\sin\alpha/\sqrt{g(R_B)}$, which can be derived from the geometry $\tan \alpha=(U^\phi U_\phi)^{1/2}/(U^r U_r)^{1/2}$~\citep{2002ApJ...566L..85B}.

Combining Eq.(\ref{eq:drdl}) and Eq.(\ref{eq:dphidl}), we find
\begin{equation}\label{eq:dphidr}
    \frac{\D\phi}{\D r}= \frac{b}{r^2}\left(\frac{fg}{1-b^2g/r^2}\right)^{1/2}\ ,
\end{equation}
Finally, by integrating Eq.(\ref{eq:dphidr}) from $R_B$ to infinity, we obtain the angle between local radial direction and the line of sight as
\begin{equation}
    \psi =\int_{R_B}^\infty \frac{\D\phi}{\D r}\D r= \int_{R_B}^\infty\frac{\sqrt{fg}}{r^2}\left(\frac{1}{b^2}-\frac{g}{r^2}\right)^{-1/2}\D r\ .
\end{equation}

We now turn to calculate the time delay, which is caused by different travel paths of the emitted photons. The time delay can be calculated by combining Eq.(\ref{eq:dtdl}) and Eq.(\ref{eq:drdl}), 
\begin{equation}
    t = \int_{R_B}^\infty \frac{\D t}{\D r} \D r =\int_{R_B}^\infty \left(\frac{f}{g}\right)^{1/2}\left(1-\frac{b^2g}{r^2}\right)^{-1/2}\D r\ .
\end{equation}
However, this integral diverges as $r\to\infty$. To avoid divergence, we define a relative time delay as~\citep{1983ApJ...274..846P,2006MNRAS.373..836P}:
\begin{equation}
    \Delta t = t(b) -t(0) =\int_{R_B}^\infty \left(\frac{f}{g}\right)^{1/2}\left[\left(1-\frac{b^2g}{r^2}\right)^{-1/2}-1\right]\D r\ .
\end{equation}
which is defined with respect to a photon emitted from the spot closest to the observer. As a result, the observer time should be corrected as $t_{\rm obs}=t+\Delta t$.

\section*{Acknowledgements}
We are thankful to W.-Z. Jiang, B. Qi, Q.-F. Xiang and the XMU neutron star group for helpful discussions. 
The work is supported by National SKA Program of China (No.~2020SKA0120300), the National Natural Science Foundation of China (Grant No.~11873040), the science research grants from the China Manned Space Project (No. CMS-CSST-2021-B11), and the Youth Innovation Fund of Xiamen (No. 3502Z20206061).

\software{PyMultiNest \citep[ \url{https://github.com/JohannesBuchner/PyMultiNest}]{2014A&A...564A.125B}.}


\end{document}